\documentclass[article,twocolumn,amsmath,floats,showpacs,nofootinbib]{revtex4}
\usepackage{graphicx}
\usepackage{textcomp}

\usepackage{hyperref}

\hypersetup{colorlinks=true}

\begin{document}

\title{Investigation of energy gap and observation of electron-phonon interaction in high-temperature superconductor $La_{1.8}Sr_{0.2}CuO_{4}$ - normal metal point contacts}
\author{I.~K.~Yanson, L.~F.~Rybal'chenko, V.~V.~Fisun, N.~L.~Bobrov, M.~A.~Obolenskii*,
Yu.~D.~Tret'yakov**, A.~R.~Kaul'**, and I.~E. ~Graboi**}
\affiliation{Physicotechnical Institute of Low Temperatures, Academy of Sciences of the Ukrainian SSR, Kharkov,\\
A. M. Gorky State University, Kharkov*,\\
and M. V. Lomonosov State University, Moscow**\\
Email address: bobrov@ilt.kharkov.ua}
\published {(\href{http://fntr.ilt.kharkov.ua/fnt/pdf/15/15-8/f15-0803r.pdf}{Fiz. Nizk. Temp.}, \textbf{15}, 803 (1989)); (Sov. J. Low Temp. Phys., \textbf{15}, 445 (1989)}
\date{\today}

\begin{abstract}Electrical characteristics $(I(eV), dV/dI(eV), d^{2}V/dI^{2}(eV))$ are studied comprehensively for pressed submicron point contacts between fresh fractures of LaSrCuO ceramic pellets and a normal metal (Cu). Point contacts (PC) with the direct-type conductivity are investigated, which are characterized by a considerable excess current due to Andreev reflection of quasiparticles at an NS boundary of high transparency. Basic requirements for point contacts suitable for PC measurements are formulated. The models of investigated point contacts permitting the interpretation of the observed characteristics are considered. The results of measurement of the energy gap reveal that LaSrCuO is a superconductor with a strong coupling since the maximum value of the ratio $2\Delta/kT\simeq 11$. It is found that superconducting properties may vary appreciably over distances of several tens of angstroms, resulting in some cases in the simultaneous emergence of two gaps corresponding to two critical temperatures. The $dV/dI(eV)$ and $d^{2}V/dI^{2}(eV)$ curves display some high-intensity singularities whose position on the eV-axis is practically independent of temperature and correlates with the phonon spectrum obtained in neutron scattering experiments. This points to a strong electron-phonon interaction in LaSrCuO.
\pacs{71.38.-k, 73.40.Jn, 73.40.Jn, 74.25.Kc, 74.45.+c, 73.40.-c, 74.20.Mn, 74.70.Ad, 74.72.-h,Dn, 74.50.+r.}

\end{abstract}
\maketitle

\section{Introduction}
Point-contact spectroscopy (PCS) is based on the investigation of nonlinearity of current-voltage characteristics of electrical point contacts (PC) with direct-type conductivity (without potential barriers). The typical size of a contact, which is determined by the maximum size of the current concentration region (we denote the contact size by $d$) must be small enough for the charge carriers "remembering" about their energy on the sides away from the contact to appear in the region with a linear size of the order of the energy relaxation length. The nonequilibrium energy distribution of charge carriers formed in this case is such that the maximum energy transfer in inelastic relaxation processes is equal to the potential difference $V$ applied across the contact. This potential difference serves as an "energy probe" in the investigation of the spectral functions of the interaction between carriers and various excitations in conductors. The PCS condition for normal metals has the form
\begin{equation}
\label{eq__1}
{d\ll min(l_{\varepsilon}, \Lambda_{\varepsilon}) ,}
\end{equation}
where $\Lambda_{\varepsilon}={(l_{i}l_{\varepsilon})}^{1/2}$ is the energy relaxation length for carriers in the dirty limit, and $l_{i}\ll l_{\varepsilon}$ ($l_{i}$ and $l_{\varepsilon}$ are the elastic and inelastic mean free paths respectively).

The chemical potential of Cooper pairs in superconductors remains constant as long as the number density of superconducting electrons differs from zero even in the presence of an electric field. The penetration depth $l_E$ of an electric field into a superconductor is determined by the relaxation of quasiparticles with an energy of the order of $\Delta$, viz., half the energy gap width. The value of  $l_E$ is normally much larger than the energy relaxation length for quasiparticles with excitation energies of the order of phonon frequencies. This allows the carriers from the two sides of an ScS contact, which "remember" their energy away from the constriction, to approach each other to within a distance smaller than the inelastic relaxation length. This determines the size of the region where the superconducting order parameter vanishes even if the contact size is larger than $\Lambda_{\varepsilon}$, and PCS is impossible in the normal state \cite{1}. A similar situation takes place in ScN contacts also provided that $d\ll \Delta_{\varepsilon}$ in the normal metal.

Thus, PCS of high-energy excitations (with energies $eV > \Delta$) is possible in ScS and ScN contacts even if the contact size does not satisfy condition \ref{eq__1}. The mechanism of emergence of singularities on current-voltage characteristics (IVC) due to inelastic interaction between the carriers and phonons or other excitations is not completely clear (see Refs. \cite{12} and \cite{13}). It is obviously associated with a partial suppression of the superconducting order parameter (reduction of the gap) for $eV \geq \hbar\omega_{i}$ ($\omega_{i}$ are the frequencies corresponding, for example, to phonons with small group velocities and high densities of states), which leads to a threshold decrease in the excess current. Experiments \cite{2,4,5} show that spectral singularities appear in the form of narrow spikes (peaks) on the $dV/dI$ characteristics at the bias $eV = \hbar\omega_{i}$. A typical feature of
phonon singularities on PC spectra is a weak temperature or magnetic field dependence of their position on the energy axis while the position of the singularities due to superconductivity degradation as a result of heating depends strongly on these parameters.

The present work reveals that the IVC of a point contact between $La_{1.8}Sr_{0.2}CuO_{4}$ and a noble metal (Cu) do indeed exhibit phonon singularities.
The positions of the peak on the $dV/dI$ curves (or the positions of singularities on the $d^{2}V/dI^{2}(V)$ curves which are close to them), and sometimes even their relative intensities, correlate with the phonon states density function for this high-temperature superconductor, which is known from neutron studies. The intensity of these peaks is much higher than the intensity of similar singularities on PC spectra of ordinary superconductors investigated earlier ($Nb, Nb_{3}Sn$ and $NbSe_{2}$), which is an indication of a very strong electron-phonon interaction (EPI) in this material. The absence of peaks on the $dV/dI(V)$ curve whose intensity would be comparable with phonon peaks at energies exceeding $k\theta_{D}$ testify to the decisive role of EPI in the superconductivity mechanism of this HTS. Preliminary results of our investigations were reported in Ref.\cite{6}.

It is well known that the energy gap $\Delta$ in the quasiparticle excitation spectrum of a superconductor leads to significant nonlinearities of IVC of ScN (or ScS) contacts with a bias $eV=\Delta$ (or 2$\Delta$). The mechanism of formation of these singularities on IVC is known only for small PC with $d<\zeta$. In the pure limit, $\zeta^{-1}=\xi^{-1}(0)+l_{i}^{-1}$ ($\xi(0)$ is the coherence length at zero temperature). A similar condition for dirty contacts has the form $d<\xi={(\xi_{0}l_{i})}^{1/2}$. In the case of HTS, these inequalities are violated in view of the extreme smallness of $\xi$. For this reason, many aspects of the mechanism of formation of gap singularities on IVC of point contacts with HTS remain unclear. The situation is also complicated by spatial nonuniformity in the properties of HTS due to their high sensitivity to the composition and structure, and is probably inherent in those layered materials whose coherence length in the direction of the $c$-axis is smaller than the size of a unit cell. Nevertheless, it will be shown that IVC of a point contact between LSCO and copper display singularities which allow us to measure the energy gap and to tract (at least qualitatively) its dependence on temperature and magnetic field. The proximity effect in this case does not lead to an anomalously strong gap suppression on the SN boundary, which could be expected proceeding from the equilibrium pattern \cite{7}. The results of gap measurements on the whole confirm the hypothesis about a strong electron-phonon interaction in this material.

Since the information on a metal investigated by the PCS method pertains to only a small region in the material with a size of the order of the contact diameter, the perfect structure and stoichiometry are insignificant beyond this region (whose size can be only a few tens of angstroms). From the methodological point of view, the difference in the investigation of ceramics, polycrystalline and monocrystalline samples consists in the number of touches required to identify a "sensitive point" having good superconducting properties in the vicinity of a contact.

\section{Experimental technique}

\subsection{Preparation of samples.}
Ceramic samples were obtained by coprecipitation from an aqueous solution containing stoichiometric amounts of $La, Sr$ and $Cu$ nitrates (X4 qualification) with a total concentration of 1 mole. In order to avoid the difficulties associated with an incomplete precipitation of cations, hot ethanol solution of oxalic acid was used. The high solubility of oxalic acid in ethanol and the laminating effect of the alcohol ensure almost complete precipitation, which was confirmed by chemical analysis of the mother liquor. The obtained precipitate was separated in a centrifuge, dried at $150{}^\circ \text{C}$ to complete sublimation of the excess of oxalic acid, and decomposed at $750{}^\circ \text{C}$ . The powder obtained after the decomposition and agglomerates was pressed and fritted for 50 h in air at $1050{}^\circ \text{C}$, and slowly cooled to $500{}^\circ \text{C}$. According to the results of $x$-ray phase analysis, the samples were one-phase compounds having the $K_{2}NiF_{4}$ structure and the lattice parameters $a$=3.777~\AA, $c$=13.241~\AA. The density of the samples amounted to 85\% of the theoretical value. According to the results of resistive measurements, the onset of the superconducting transitions was at $38~K$, the transition width varying within several degrees for different samples.

\subsection{Creation of point contacts and measurement of their characteristics.}
A point contact was formed at the point of contact between sharp edges of a normal ($Cu$) and superconducting ($LaSrCuO$) electrodes having an elongated shape and a size of a few millimeters (see inset in Fig. \ref{Fig2}). The copper electrod was cut on an electric spark machine from a bulk ingot and polished electrolytically at the final stage in an acid mixture whose composition is described in Ref. \cite{5}. The S-electrode was cleaved with a scalpel from an original pellet and cemented to a damper made of beryllium bronze with the help of a silver paste ensuring a reliable electric contact. Then the two electrodes were mounted in a special device permitting their relative displacement directly in the cryostat, which made it possible to renew the point of contact many times during a single measuring cycle and to control the force of pressure. The time interval between the chopping of the S-sample and its mounting in the cryostat did not exceed 10 min.

\begin{figure}[]
\includegraphics[width=8cm,angle=0]{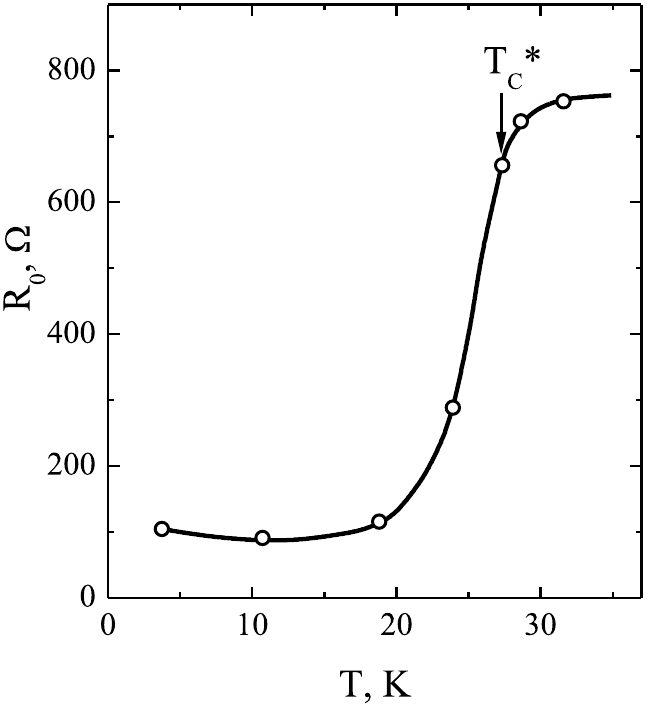}
\caption[]{Temperature dependence of resistance at $V=0$ for contact No.1.}
\label{Fig1}
\end{figure}

\begin{figure}[]
\includegraphics[width=8cm,angle=0]{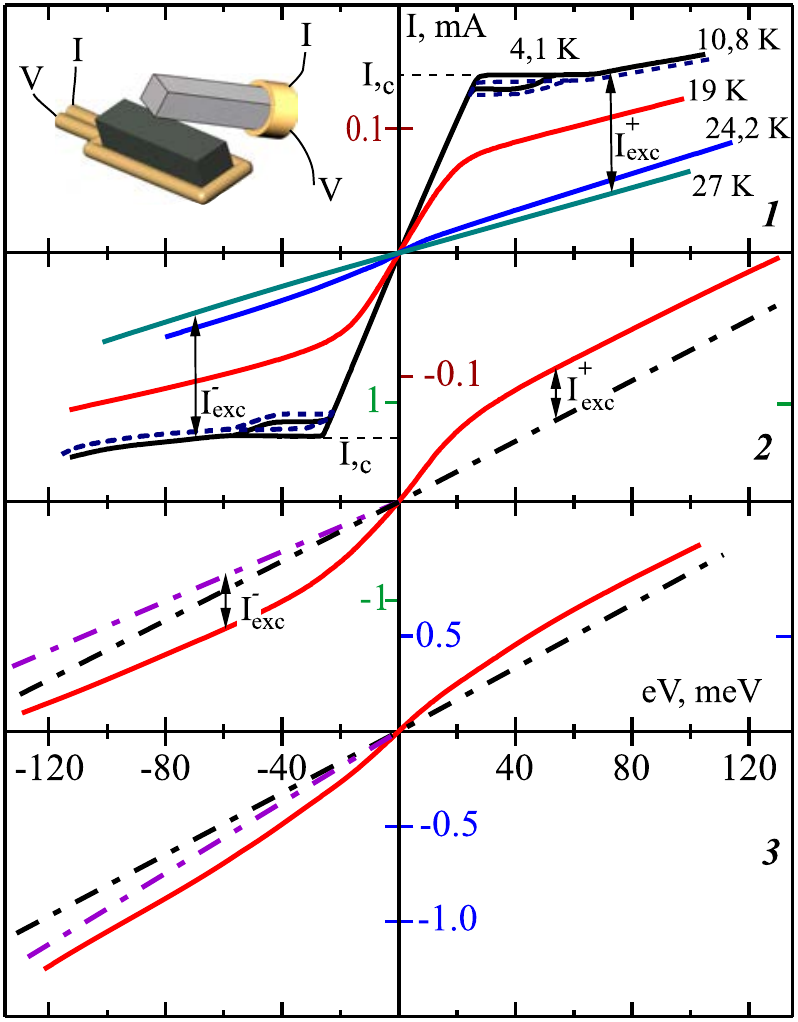}
\caption[]{IVC for contacts Nos.1, 2, and 3 (the numbers on the curves correspond to the numbers of contacts in Table \ref{tab1}). The IVC for contact No.1 are taken at different temperatures; IVC for contacts Nos.2 and 3 are taken at $T=1.57$ and $4.4~K$, respectively. The dot-and-dash straight lines illustrate the method of determining $R_N$. The inset shows the geometry of the experiment.}
\label{Fig2}
\end{figure}

The quality of a point contact was estimated from the form of IVC and its derivatives as well as from the temperature dependence of resistivity at zero bias (Fig. \ref{Fig1}).

Firstly, it is necessary that the contact be formed in the region of emergence of the superconducting phase with a high $T_{c}^{*}$ on the surface of a sample. For this reason, the temperature of the superconducting transition in the material in the immediate proximity of the contact was measured in most cases. This temperature $T_{c}^{*}$ is normally slightly lower than the temperature $T_c$ of the resistive transition in the bulk sample. It is determined either from a sharp drop in the contact resistance at zero bias with lowering temperature (Fig. \ref{Fig1}) or from the emergence of a minimum near $V=0$ on the $dV/dI(V)$ curve (see, e.g., Fig. \ref{Fig5}).
The more clearly these properties are manifested, the closer the superconducting region is to the physical boundary between copper and ceramic, and the larger the volume of the contact occupied by it.

\begin{table*}[]

\caption []{Parameters of Point Contacts}

\begin{tabular}{|c|c|c|c|c|c|c|c|c|c|c|c|c|c|c|}\hline
No, of & \multicolumn{2}{|c|}{$R_{N},\Omega$}& $R_{0}$, & $T_{c}^{*},$ & \multicolumn{2}{|c|}{$\Delta_{1},meV$} & \multicolumn{2}{|c|}{$\Delta_{2},meV$} & \multicolumn{2}{|c|}{$\Delta_{m},meV$} & ${2{{{\bar{\Delta }}}_{1}}}/{kT_{c}^{*}}\;$&${2{{{\bar{\Delta }}}_{2}}}/{kT_{c}^{*}}\;$ &${2{{{\bar{\Delta }}}_{m}}}/{kT_{c}^{*}}\;$& Remark \\ \cline{2-3} \cline{6-11}
contact & + & $-$ & $\Omega$ & $K$ &\  + \ & $-$ & + & $-$ & \ \ + \ & $-$ & & & & \\ \hline
1& 1110 & 1546 & 91 & 27.4 & 6.8 & 6.1& 13.2 & 12.6& -&-&5.46&10.93&-&-\\ \hline
2&74&90&32&30&-&-&12.3&13 & \multicolumn{2}{|c|}{$2.5\div 13$}&-&9.79&$0.97\div5.03$&-\\ \hline
3&126&107&92&31&-&-&-&-&9.8&8.6&-&-&6.89&-\\ \hline
4&159&153&-&19.6&-&-&-&-&7.4&7&-&-&8.53&-\\ \hline
5&1935&1962&715&20&4.2&3.6&7.7&7.2&-&-&4.53&8.65&-&-\\ \hline
5&2185&-&830&20&-&-&-&-&6.4&6.0&-&-&7.20& after application\\
&&&&&&&&&&&&&&of magnetic field\\ \hline
6&118&118&116&31&-&-&\multicolumn{2}{|c|}{$10\div11$}&\multicolumn{2}{|c|}{$4\div11$}&-&$3.74\div4.12$&$1.54\div4.12$&-\\ \hline

\end{tabular}
\label{tab1}
\end{table*}
Second, it is important that the conductivity must be of the metal type at a large bias ($eV>\Delta$), i.e., it must decrease with increasing voltage.
In this case we assume that the voltage dependence $R_{D}(V)$ of the differential resistance is in qualitative agreement with the temperature dependence of the resistivity of the bulk sample as in PCS of normal metals, while the metal-type behavior of $R_{D}(V)$ for $eV>\Delta$ corresponds to a linear dependence $\rho(T)$ of perfect LSCO single crystals for $T>T_{c}$. The deviations from the periodicity of the crystalline structure lead to a violation of the coherent ("band") motion of charge carriers and necessitate the introduction of a tunnel or jump-like mechanism of conductivity with a negative temperature coefficient.
It should be noted that most works on point-contact spectroscopy of HTS were carried out with the tunnel mode corresponding to $R_{D}$ decreasing with voltage.

Third, a high (of the order of $\Delta/R_{0}$) and weakly dependent excess current $I_{exc}$ must be present on IVC in the superconducting state (Fig.\ref{Fig2}, IVC for contact No.2). The large magnitude of the excess current is an indication of the absence of potential barriers and normal leakage currents shunting the contact or, in other words, of the fact that the most part of the contact region is filled with the superconducting phase. The constancy of the excess current points towards the absence of a noticeable heating of the contact region.

Fourth, the resistance of a point contact must be as high as possible to correspond to a small size $d$. In this case, the probability of complete filling of the contact region with a homogeneous superconducting phase has a maximum value. In experiments, resistances of the order of several kiloohms were attained (see Table \ref{tab1}, contact No.5), which corresponds to a contact size of the order of a few tens of angstroms.

Fifth, the contact diameter $d$ determined from the resistance $R_0$ at zero bias according to the Sharvin formula
\begin{equation}
\label{eq__2}
{d={{\left( \frac{16{{(\rho l)}^{Cu}}}{3\pi {{R}_{0}}} \right)}^{1/2}}\quad ,}
\end{equation}
must correspond to its value calculated by Maxwell's formula
\begin{equation}
\label{eq__3}
{d=(1/2)(\rho_{\text{LSCO}}/R_{N}) ,}
\end{equation}
where $R_N$ is the resistance in the normal state, which is determined by the superconducting bank almost completely. This condition presumes that a contact can be modeled by a circular aperture in a thin partition which divides the normal and the superconducting banks and is impenetrable for electrons.
It is also assumed that the superconducting phase comes close to the interface (sharp boundary), filling the contact region completely. This ideal pattern is not confirmed in experiments as a rule.
The departures from this pattern are so large that the above condition is not satisfied even to within an order of magnitude.

Finally (the sixth criterion) phonon peaks on IVC derivatives should not be displaced appreciably on the energy axis upon a change in temperature or magnetic field, while their intensity and width may considerably depend on these parameters. On the contrary, stray spikes due to the superconductivity degradation along the current lines on the $eV$-axis towards zero with increasing $T$ or $H$, their intensity and sharpness remaining unchanged up to temperatures close to $T_c$.
\begin{figure}[]
\includegraphics[width=7.5cm,angle=0]{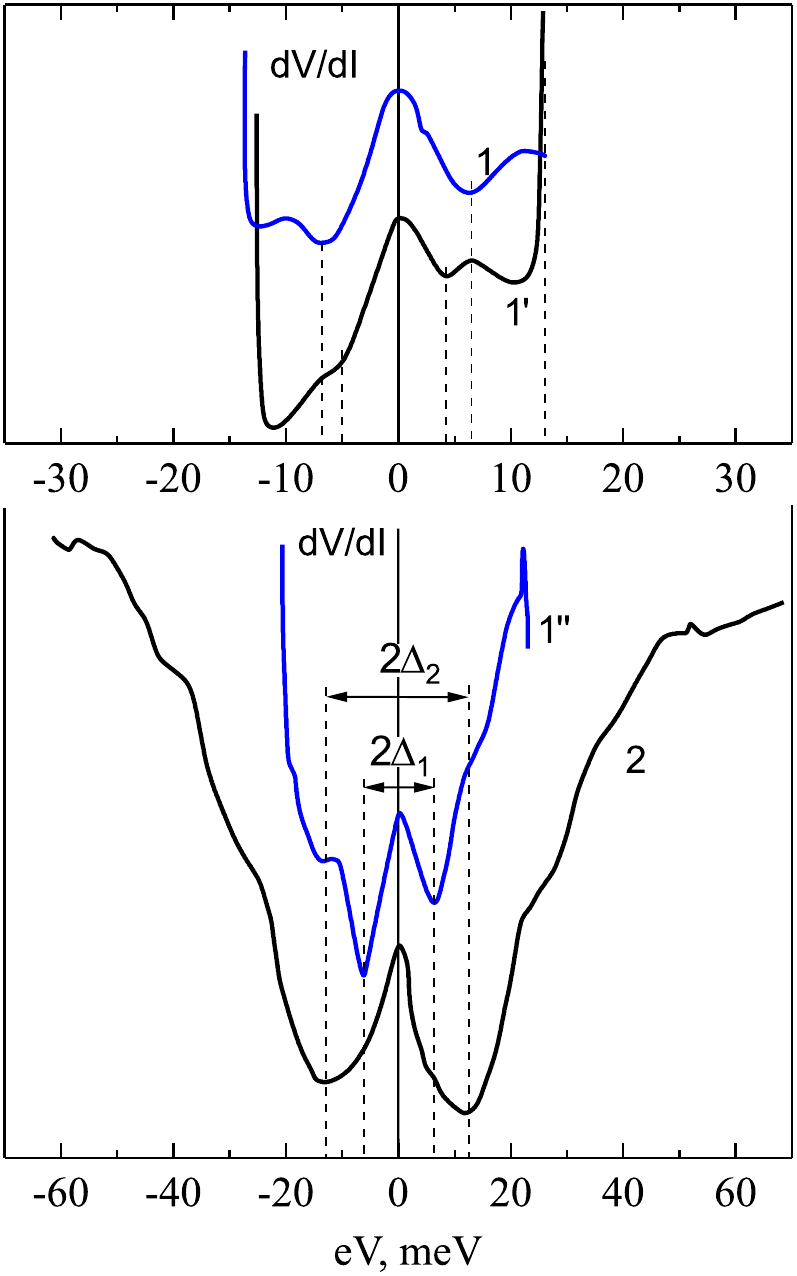}
\caption[]{The first derivatives of IVC for contacts Nos.1 and 2. Curves 1 and 1' were recorded at 4.1 and $10.8~K$, respectively, while curve 1" was recorded in another series of experiments at $1.6~K$ (it repeats curve 1 in Fig. \ref{Fig15}b on a reduced scale for $V$). The vertical dashed lines project the positions of $dV/dI$ peaks onto the $eV$-axis.}
\label{Fig3}
\end{figure}

\begin{figure}[]
\includegraphics[width=7cm,angle=0]{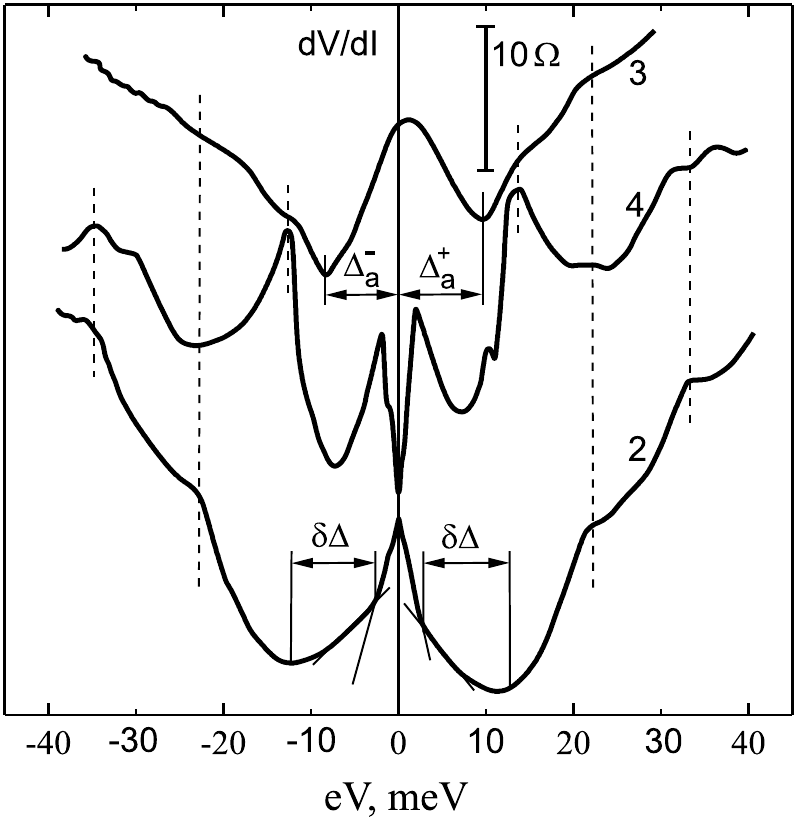}
\caption[]{Differential resistance versus bias voltage for contacts Nos.2-4 (curve 2 repeats curve 2 in Fig.\ref{Fig3} on a magnified scale); curves 3 and 4 were recorded at 5.0 and $4.2~K$, respectively. Dashed vertical lines project phonon singularities onto the energy axis. The methods of determining $\Delta^{\pm}$ (curve 3) and the interval $\delta\Delta$ over which the gap for contact No.2 is "spread" are indicated.}
\label{Fig4}
\end{figure}

Current-voltage characteristics and their derivatives were measured on a point-contact spectrometer operating on the modulation principle. The signals proportional to the first
 ($V_{1}(eV) \propto dV/dI$) and second ($V_{2}(eV) \propto d^{2}V/dI^{2}$) harmonics of the modulating voltage were registered (after amplification and synchronous detecting) on an x-y recorder. The initial component of the modulating voltage was compensated while measuring $V_{1}(eV)$ with the help of a bridge circuit. Relative nonlinearities of IVC were registered with an accuracy not lower than $10^{-3}$ and $10^{-5}$ for the first and second harmonics channels, respectively. Measurements were made in the temperature interval between 1.5 and 300~$K$ in a special cryostat formed by two Dewar flasks embedded one into the other. Cryogen was delivered into the inner cryostat through a capillary in a vacuum jacket. The temperature was controlled either by varying the pumping rate ($T<4.2~K$ for measurements in liquid helium) or by using a resistive heater ($T>4.2~K$ for measurements in vapors). In the latter case, the pressure in the outer cryostat was kept at 30-40~mmHg higher than in the inner cryostat. The magnetic field was produced by a superconducting solenoid and could be varied between 0 and 50~$kOe$.

\section{PECULIARITIES OF IVC AND THEIR DERIVATIVES}

Let us describe some peculiarities of IVC for LSCO/Cu point contacts which do not refer directly to phonon or gap spectroscopy. The correct interpretation of these peculiarities is important for appropriate interpretation of PC spectra.
\subsection{IVC asymmetry}
Figure \ref{Fig2} shows IVC for three point contacts. The first derivatives for these contacts and contact No.4 (see Table \ref{tab1}) are represented in Figs.\ref{Fig3} and \ref{Fig4}.

 These characteristics are weakly asymmetric. The following parameters are different for positive and negative $eV$ of equal magnitudes: (1) asymptotic resistances $R_{N}^{\pm}$ determined by the slopes of the straight lines parallel to excess current; (2) the values $I_{exc}^{\pm}$ of excess current for the bias of the same magnitude; (3) the depth of the gap minima of $dV/dI$, whose asymmetry usually correlates with the asymmetry in $R_N$, and (4) the positions of $R_D$ minima from which the energy gap $\Delta^{\pm}$ is counted. It can be seen that there is a correlation between $I_{exc}^{\pm}$ and $R_{N}^{\pm}$, while there is no correlation between these parameters and $\Delta^{\pm}$. The asymmetry is apparently due to the dependence of the height of the small potential barrier at the interface on the polarity of the applied voltage.
Since the transparency of the barrier for the contacts under investigation is close to unity, the asymmetry is very small. Tunnel contacts exhibit, as a rule, much stronger asymmetry in IVC and its derivatives \cite{8}. It should be noted that the excess current remains constant up to biases exceeding the energy gap by more than an order of magnitude. This is an indication of the absence of a noticeable heating of HTS in the contact region. The IVC for contacts Nos.2 and 3 in Fig.\ref{Fig2} has a negative curvature ($d^{2}I/dV^{2}<0$), which points towards the metal type of the resistance offered to quasiparticle excitations with energies exceeding $\Delta$ at $T \ll T_{c}^{*}$.
The curvature is positive for the high-resistance contact No.1 in spite of the large excess current.
Let us assess the lower limit of the diameter of this contact, assuming that the electron flight in the copper bank is ballistic (formula (\ref{eq__2})). For $R_{0}=83~\Omega$
 and $(\rho l)^{Cu}=0.66 \cdot 10^{-11} \Omega\cdot cm^{2}$ we obtain $d\geq 37 \AA$. On the other hand, by assuming that the motion of electrons in the normal HTS is diffusive ($l_{i}\ll d$), we find from formula (\ref{eq__3}) the value $d<67~\AA$  for the upper limit by putting $l_{\text{LSCO}}\leq10^{-3}~\Omega\cdot cm^{2}$, $R_{N}=750~\Omega$. Thus, the fifth quality criterion is satisfied for this contact. This criterion is obviously violated for contacts Nos.2 and 3; consequently, their configuration differs considerably from the ideal model.

\begin{figure}[]
\includegraphics[width=7cm,angle=0]{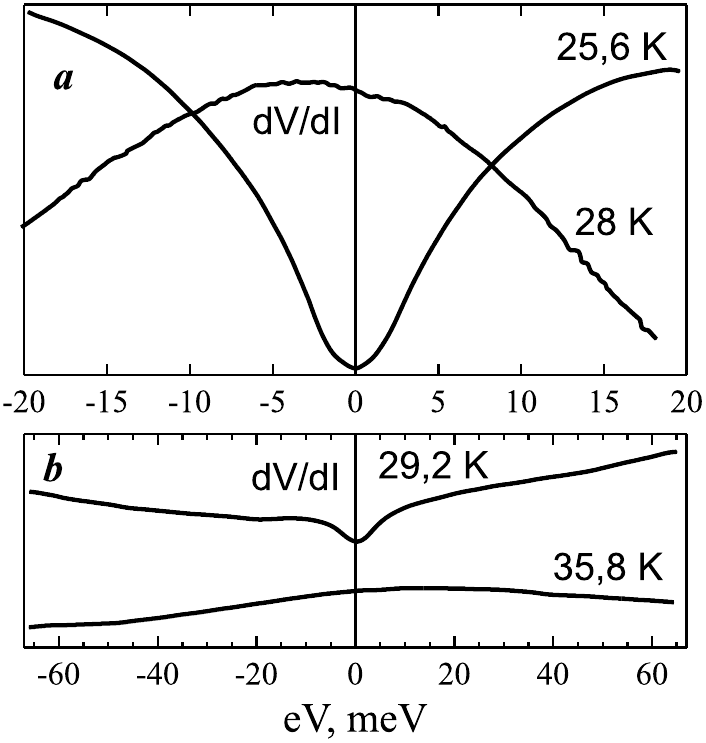}
\caption[]{"Metal-insulator" transitions for the first derivatives of IVC for contacts No. 1(a) and 6(b) at different temperatures embracing the superconducting transition temperature. The ordinate scale is in arbitrary units, and the curves are displaced arbitrarily along the ordinate axis.}
\label{Fig5}
\end{figure}

\begin{figure*}[]
\includegraphics[width=15cm,angle=0]{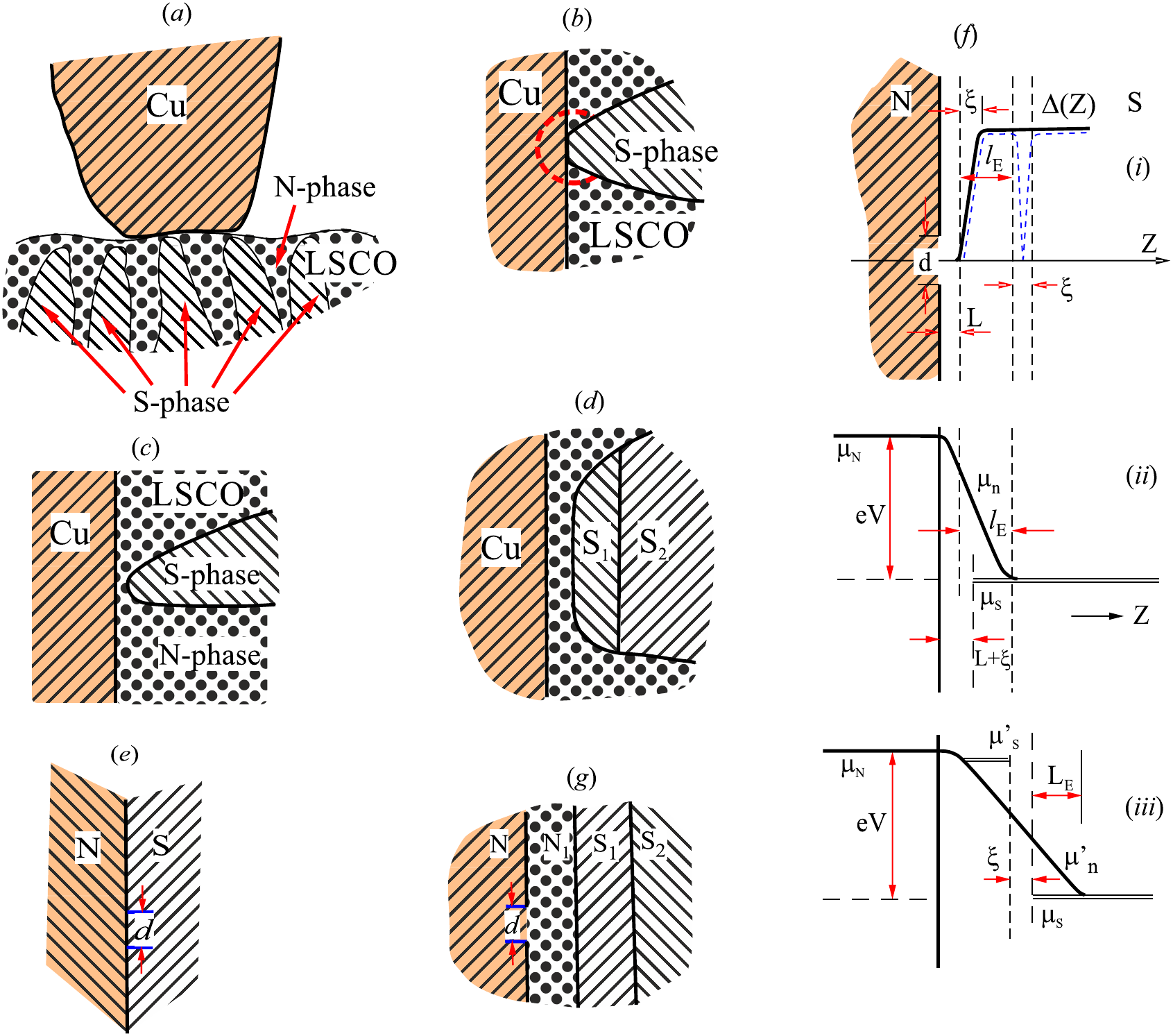}
\caption[]{Schematic diagram of point contacts between HTS ceramics and copper: structure (a), ScN and SNcN type contacts modeling an ideal and real point contact (b,c); schematic of a real point contact with alternating real ($N_1$ ) layer and two different HTS phases ($S_1$ and $S_2$) (d, g); an aperture in an impenetrable partition modeling an ideal ScN contact (e); models of a real contact of the ScNS type (f); the coordinate dependence of energy gap in equilibrium (solid curve) and in the current mode (dashed lines) during the formation of PSC (i), and the coordinate dependence of chemical potential $\mu_N$ of quasiparticles and $\mu_S$ of Cooper pairs for $eV>\Delta$ in the absence of PSC (ii) (the same parameters are shown on (iii) diagram for the case of PSC formation).}
\label{Fig6}
\end{figure*}
\subsection {Zero-point singularities of differential resistance}
In the vicinity of $V=0$, the differential resistance normally has a peak corresponding to a small tunnel component of the current (curve 3 in Fig.\ref{Fig4}). In the case of a sharp SN boundary,
a potential barrier can emerge due to different electronic characteristics ($p_F, v_F$) of the metals in contact. If there are no centers destroying Cooper pairs at the boundary, the proximity effect "pulling" superconductivity into the normal metal must take place. The role of the normal metal can be played both by $Cu$ (in the case of a sharp boundary) and by the surface LSCO layer depleted in oxygen.
In both cases, a narrow minimum of $dV/dI$ is observed at $eV=0$ (see, for example, contact No.4 in Fig.\ref{Fig4}), whose depth increases with lowering temperature.
A small current through the contact destroys this superconductivity, and a gap that is not subjected to the proximity effect will be observed in both cases. Similar narrow minima of $dV/dI(0)$, strongly depending on temperature and magnetic field, were also observed for ScN contacts between $Ta$ and $Cu$ \cite{9}.

A zero-point anomaly in $R_D$ with a narrow minimum is apparently observed in contacts with a Schottky tunnel barrier formed due to spatial separation of charges at the boundary. Figure \ref{Fig4} (curve 2) shows that this barrier has an insignificant height, and we go over the conventional form of $dV/dI(V)$ with gap minima even at a comparatively small bias.
\subsection {"Metal-insulator" transition on $R_D(V)$ curves for $T=T_c^*$}

It is well known that the $R_D(V)$ curve for a point contact formed by traditional superconductors at $eV>\Delta$ differs only slightly from the corresponding dependence in the normal state on the one hand and is qualitatively similar to the temperature dependence of the resistivity of the bulk material above $T_c$ on the other hand. In point contacts containing HTS, a different situation, conventionally referred to as the "metal-insulator transition", often takes place at $T=T_c^*$. Figure \ref{Fig5}
shows differential resistances as functions of the bias voltage at different temperatures embracing the superconducting transition temperature of the contact. It can be seen that at $T\leq 27.4~K$, the emergence of a minimum of $dV/dI$ is accompanied by a change in the sign of the second derivative of IVC for large $eV$ at zero bias, indicating the onset of superconductivity. The semiconducting (or tunnel) type of the $R_D(V)$ dependence in the normal state changes for the metal-type dependence for $T<T_c^*$. (This effect can be clearly seen on the temperature dependences of IVC in Figs.\ref{Fig10}, \ref{Fig12}, and \ref{Fig13}).

A simple explanation of this effect can be given on the basis of possible nonuniformity in the properties of the superconductor in the contact region. Let us suppose that defects of structure and composition of HTS are localized on the periphery of the contact. However, it is these defects that determine the resistance of the contact and its semiconductor-type temperature dependence in the normal state. In the superconducting state, the resistance of this region is shunted by supercurrent provided that the electric field penetration depth $l_F$ is smaller than the contact diameter \emph{d}, and the sign of the derivative $d^2V/dI^2$ changes for $eV>\Delta$ since it is now larger part of the constriction. In this case,
however, the differential resistance in the superconducting state at a large bias must be considerably smaller than the resistance in the normal state. Unfortunately, this was not confirmed in experiments. For example, the "metal-insulator transition" is observed for contact No.1 in Fig.\ref{Fig13} at $T_c^*=27.4~K$, while the family of IVC recorded for this contact at different temperatures indicates that the differential resistance at a large bias practically remains unchanged upon a transition through $T_c$.

There is also a contradiction in the estimate of $l_E$ which, if the above arguments are correct, should not exceed a few tens of angstroms (it should be recalled that the estimation of the size of contact No.1 gives $37~\AA<d<67~\AA$). On the other hand, differential resistances in experiments at larger bias are approximately equal also for contacts with lower resistance and a size of the order of hundred angstroms and higher.

In our opinion, these contradictions can be removed by assuming the possibility of formation of phase slip centers (PSC) along the current lines in HTS. The experimental facts described in the next section also point towards the existence of PSC.
\begin{figure}[]
\includegraphics[width=7cm,angle=0]{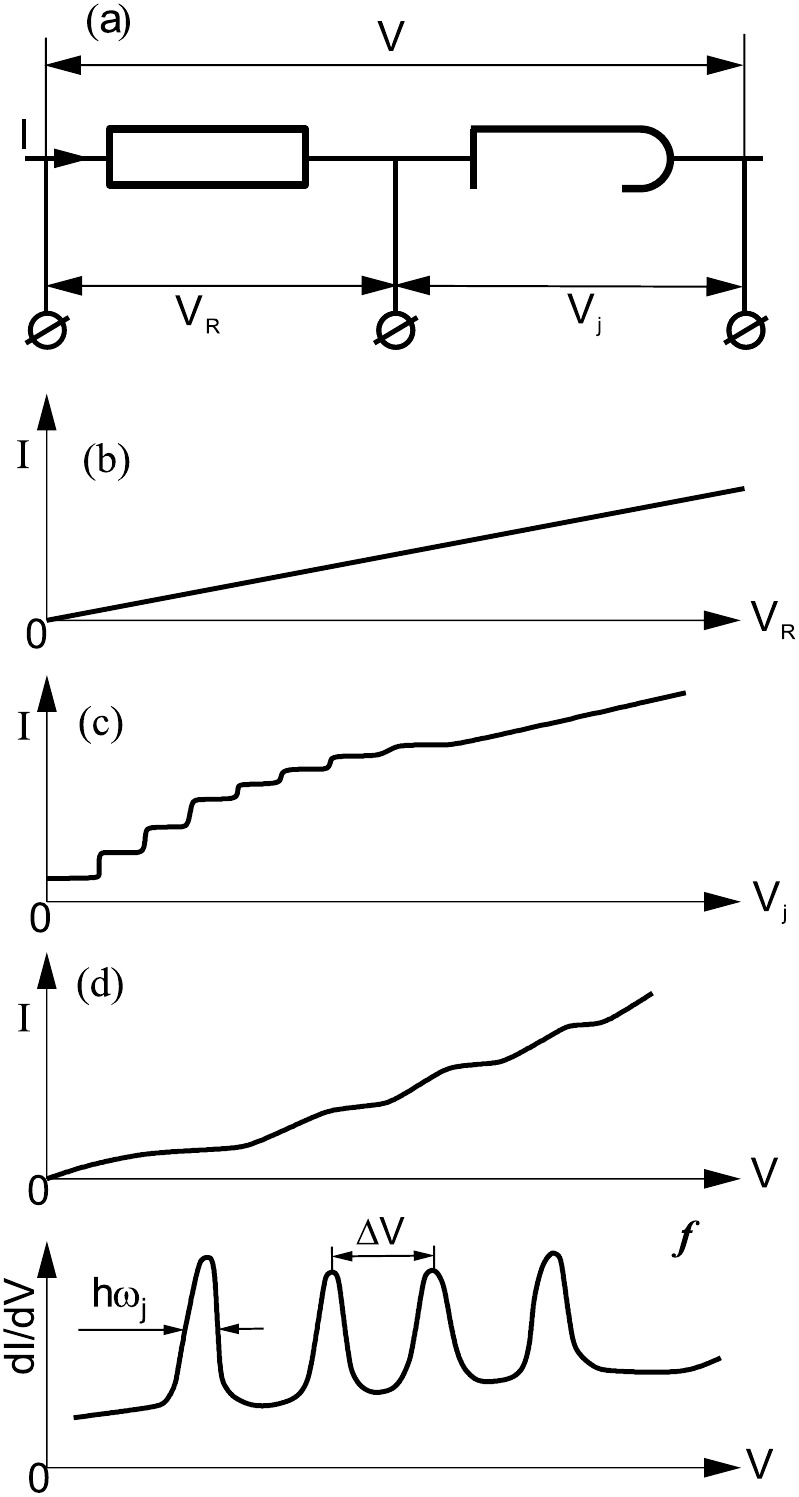}
\caption[]{Equivalent circuit diagram and IVC of an ScNS contact with PSC (models \emph{e} and \emph{f} in Fig.\ref{Fig6}) exposed to microwave radiation.}
\label{Fig7}
\end{figure}

\begin{figure}[]
\includegraphics[width=7cm,angle=0]{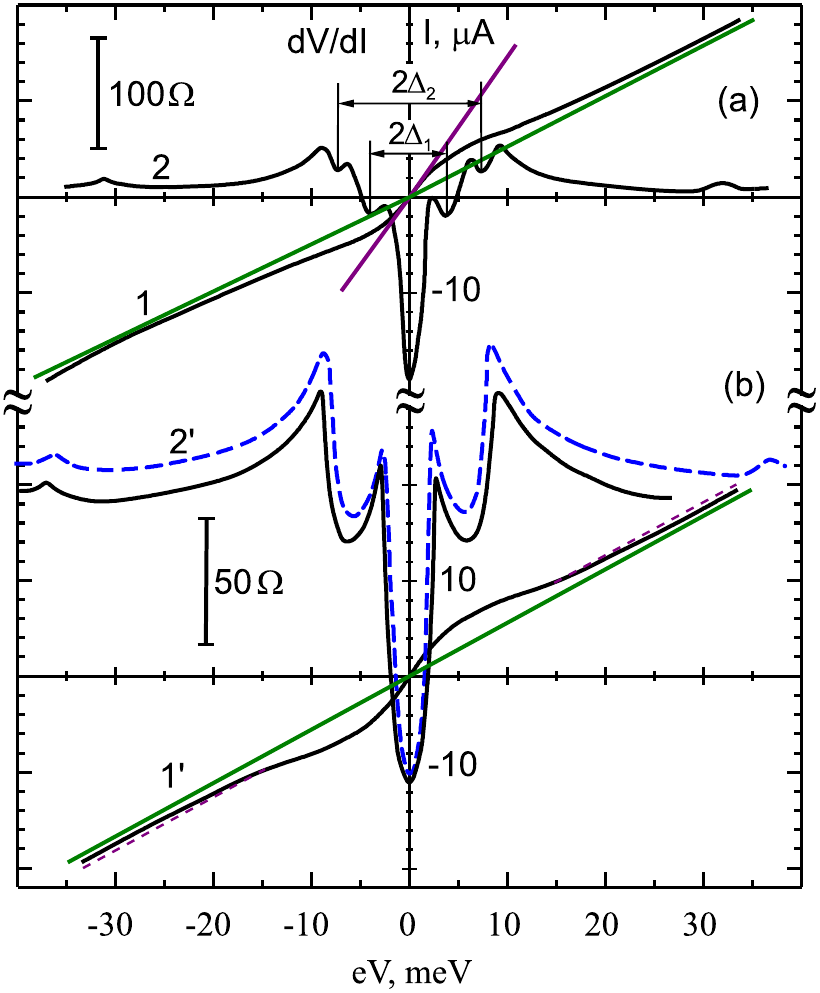}
\caption[]{IVC and their derivatives for contact No.5 before (a) and after (b) the action of microwave radiation (solid curves). Dashed curves illustrate the variation of $dV/dI(V)$ and IVC under the action of microwave radiation at $T=4.1~K$; thin lines make it possible to determine $I_{exc}$, $R_N$, $R_0$ and $\Delta_{1,2}$.}
\label{Fig8}
\end{figure}

\begin{figure}[]
\includegraphics[width=7cm,angle=0]{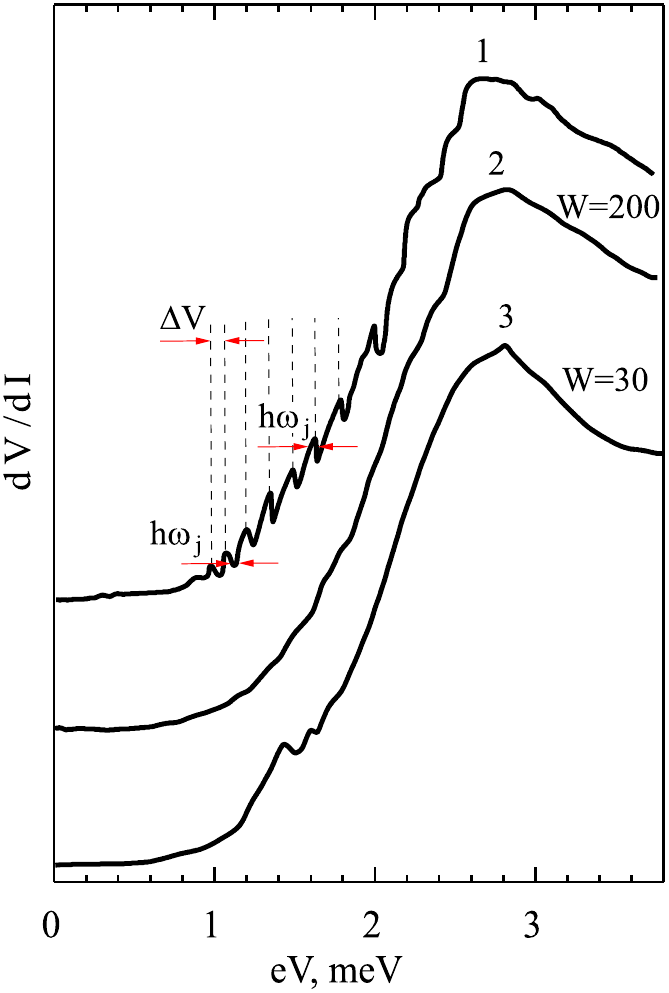}
\caption[]{$R_D(V)$ dependence for different intensities of microwave radiation: 1-$dV/dI$ structure reflecting Josephson steps on
IVC of an SNS junction formed by PSC in HTS electrode upon micro-wave irradiation; curve 2 corresponds to the maximum intensity of microwave radiation, which is almost an order of magnitude higher than the microwave intensity level for curve 1; curve 3 is the same as curve 1, but recorded after the action of high-intensity microwave radiation.}
\label{Fig9}
\end{figure}

\begin{figure}[]
\includegraphics[width=7.5cm,angle=0]{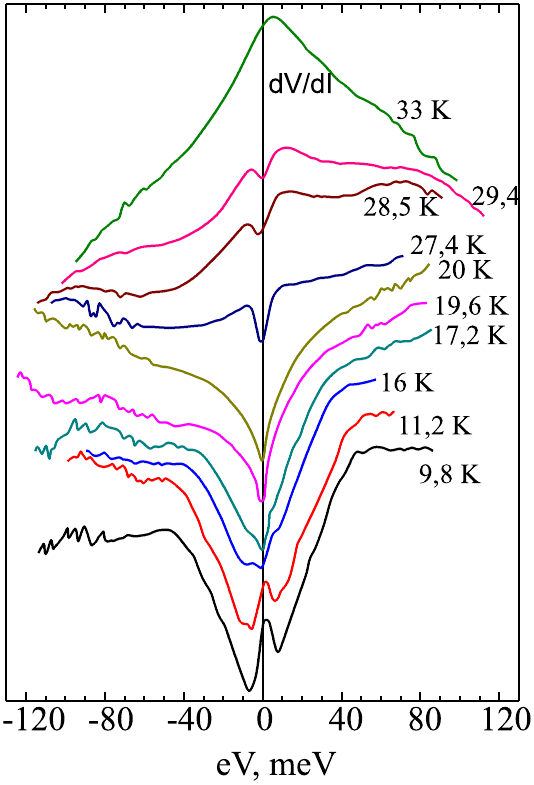}
\caption[]{$dV/dI(V)$ curves for contact No.3 at different temperatures.}
\label{Fig10}
\end{figure}

\begin{figure}[]
\includegraphics[width=7.5cm,angle=0]{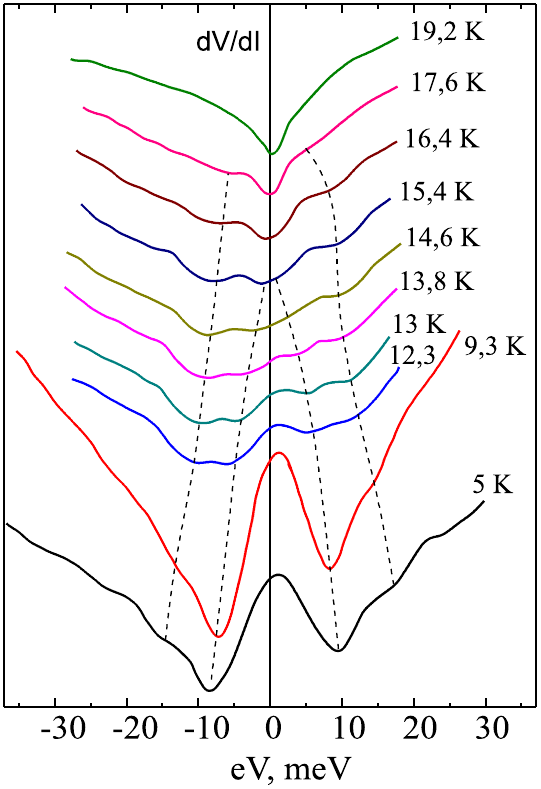}
\caption[]{$dV/dI(V)$ dependence for contact No.3 with a smaller temperature step in the vicinity of $T=T_{c1}$ where $\Delta_1$ vanishes.}
\label{Fig11}
\end{figure}

\begin{figure}[]
\includegraphics[width=7cm,angle=0]{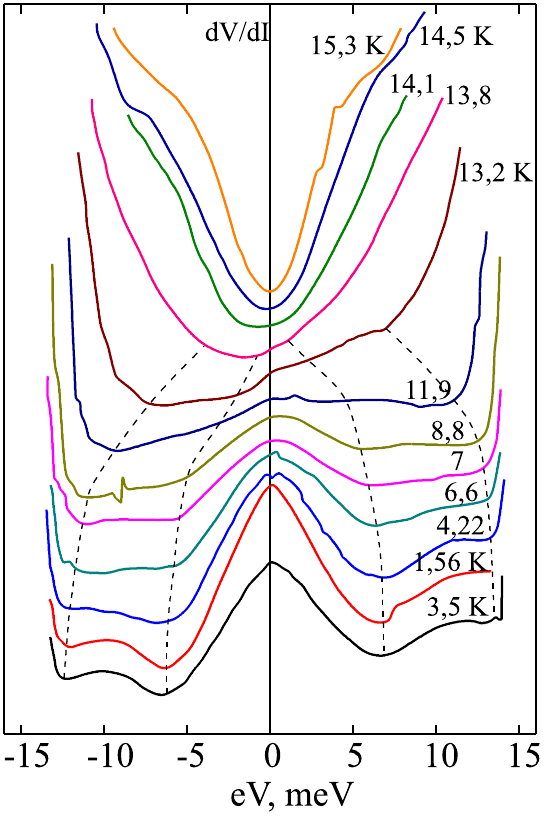}
\caption[]{$dV/dI(V)$ dependence for contact No.1 at different temperatures.}
\label{Fig12}
\end{figure}

\begin{figure}[]
\includegraphics[width=7cm,angle=0]{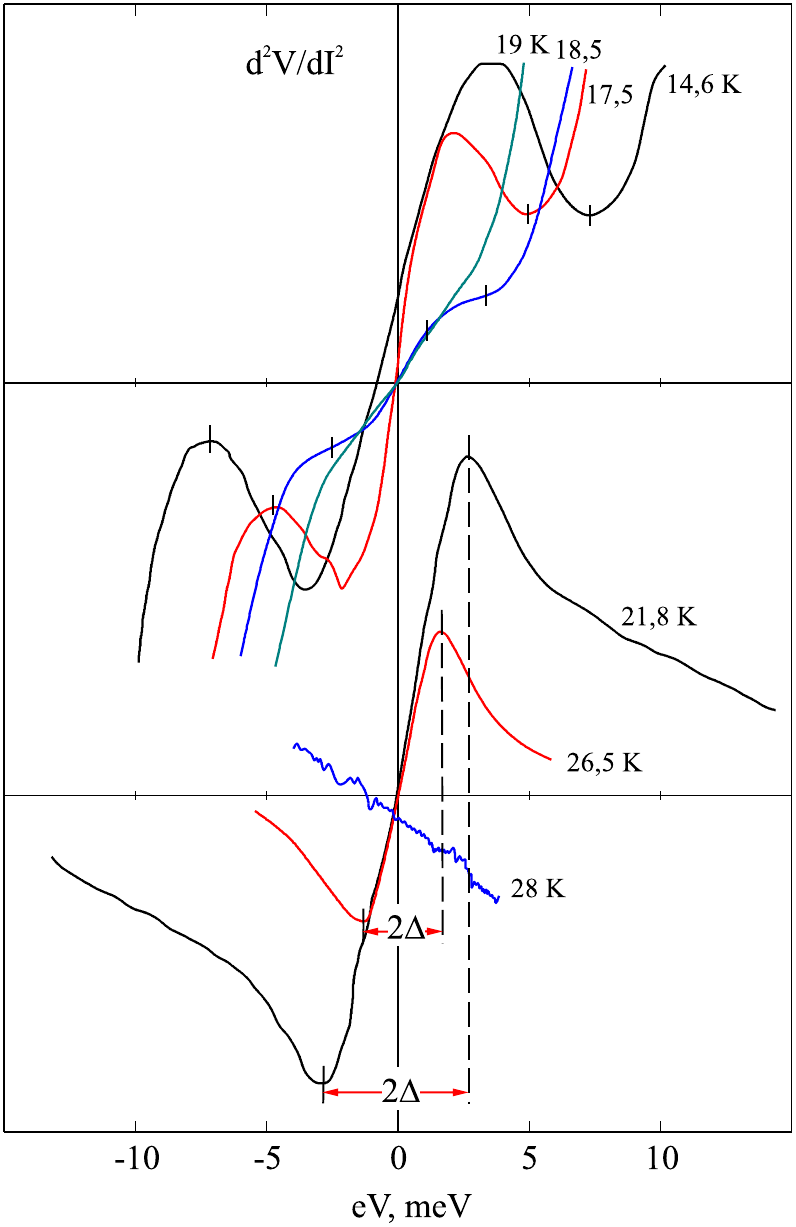}
\caption[]{Second derivative of IVC for contact No.1 at different temperatures near $T_c^*$.}
\label{Fig13}
\end{figure}

\section{CONTACT MODELS}
Although the mechanical contact between the electrodes is formed over a large area, thus ensuring the stability of the construction, the electric contact emerges only in small regions free of thin insulating (oxide and other) films. The surface layer of HTS consists mainly of a poorly conducting or insulating material. For this reason, we have to look for rare regions where deep superconducting layers emerge at the surface (Fig.\ref{Fig6}a) for the formation of a contact. In some cases, a superconducting crystallite is in direct contact with the pure surface of the normal metal (Fig.\ref{Fig6}b), but the situation when the superconducting phase is quite close to the physical boundary and yet does not touch it is more probable (Fig.\ref{Fig6}c). The two phases are separated by a thin layer of superconducting metal phase of HTS, whose thickness is small in comparison with the inelastic mean free path. A more complicated case of the superconducting phase with a nonuniform composition is also possible, when the HTS surface is formed by a sequence of the normal phase, the superconducting phase $S_1$ with a lower critical temperature $T_{c1}$ , and deeper layers of the superconductor with the critical temperature equal to the transition temperature of the bulk sample (Fig.\ref{Fig6}d). The corresponding models of \emph{e}, \emph{f}, and \emph{g} contacts are also presented in Fig.\ref{Fig6}. Potential barriers with transparency of the order of 0.1-1.0 can be present at the boundaries. It should be noted that the proximity effect described in the previous section is observed for the model \emph{e}. It can be expected that the fifth criterion of contact quality will be fulfilled for this model. On the contrary, for models f and g the influence of the proximity effect on the electron made of the normal metal is insignificant in view of the smallness of the coherence length in HTS, and the differential resistance of the contact for small and large bias voltages must be of the same order of magnitude as in the normal state.

Let us now consider the energy level diagram for the model f of a point contact in the current state at low temperatures. As long as the energy of quasiparticles impinging on the superconductor from the normal metal
 is smaller than the equilibrium value $\Delta$ of the gap, the quasiparticles undergo an Andreev-type reflection at the jump $\Delta$. The resistance in this bias region ($eV<\Delta$) is equal to the sum of Sharvin's resistances on the side of the normal metal and the resistance of a thin layer of nonsuperconducting HTS, which apparently plays a decisive role. As soon as $eV$ becomes higher than $\Delta$, quasiparticle excitations with an energy exceeding the gap will be responsible for the transport of excess current over the maximum possible superconducting current density to the depth $l_E$ into the bulk of the superconductor. It is natural to assume that the maximum value $\Delta_{max}$ of the gap corresponds to the drift of carriers along well-conducting basal planes which shunt the spreading of the current in the transverse direction. If the "easy" lines for the superconducting current are interrupted by defects, phase slip centers (lines or surfaces) emerge in the vicinity of these lines, at which the electrochemical potential of pairs undergoes a jump. Such a PSC is actually a Josephson SNS junction.
The threshold of quasiparticle current through it is $eV=2\Delta$, while the frequency of Josephson's alternating current is $2ev/\hbar$ as usual. The emergence of such PSC leads to additional mechanisms of the manifestation of phase singularities on IVC as well as to a nonequilibrium suppression of the excess current at phonon energies, which forms the basis of EPI spectroscopy for these materials. It should be noted that the behavior of the chemical potential of pairs and quasiparticles near a sharp Sn boundary for $eV>\Delta$ is similar to that of half PSC but without the Josephson effect.

The emergence of the PSC can be proved by analyzing the interaction between the Josephson current and an external electromagnetic radiation.
An equivalent circuit diagram for such a PSC consists of a resistance $R_1$ connected in series with a Josephson element (Fig.\ref{Fig7}a). The current-voltage characteristics of these elements (in a given current regime) are shown schematically in Fig.\ref{Fig7}b, and c; the resultant IVC is given in Fig.\ref{Fig7}d. Its first derivative contains a series of peaks whose width is of the order of a Josephson quantum $\hbar\omega_j$, while the peak separation is determined by the height of Shapiro steps in current and by the values of $R_1$ The latter may generally depend on \emph{V}, which will lead to a change in the separation between $R_D$ peaks.

Figures \ref{Fig8} and \ref{Fig9} present IVC and $R_D(V)$ for contact No.5 (see Table\ref{tab1}) exposed to microwave radiation and in the absence of it. The differences between curves 2 and 2' in Fig.\ref{Fig8} are apparently due to the magnetic flux trapping as a result of suppression of superconductivity near PSC by a high-intensity microwave field. After the action of this radiation, a single broad minimum is observed instead of two minima of 3.8 and 7.4~$meV$ on $R_D(V)$ in this interval of bias voltage. The spikes appear on the $R_D(V)$ dependence not at zero voltage; in
other words, there is a certain threshold in current (curve 1 in Fig.\ref{Fig9}). The separation between the steps considerably exceeds $\hbar\omega_j$, increasing with the step number as a result of the increase in $R_1$ on this bias interval. The width of the peaks is approximately equal to a Josephson quantum. High intensities of microwave radiation smear the clear-cut spatial structure of PSC due to nonequilibrium effects (curve 2). No return to the previous pattern at low-intensity microwave radiation is observed (curve 3), which is an indication of the magnetic flux trapping in the contact region, although the effect is reproduced qualitatively.

It is appropriate to consider a few remarks of general nature. First, if we exclude the artefacts due to a possible partial transport of the superconducting material to the surface of the normal electrode, in the experiments where Josephson effects were observed on ScN contacts the latter are obviously due to the formation of a phase slip surface in the bulk of the superconducting electrode in the vicinity of the SN boundary, this concerns especially the superconducting materials with a short coherence length, viz., HTS, compounds with heavy fermions, etc. The threshold in constant current is not necessary in this case since PSC can be formed by the microwave radiation itself in the absence of a transport current also \cite{10}.

Second, the oscillating singularities on IVC for contacts in a nonequilibrium state \cite{9,11} (see also Refs.\cite{12,13}) are obviously due to the entrance and motion of vortices (or current tubes) in the contact region and due to the formation of a phase slip surface. It should be noted that such a point of view differs from the interpretation of the results obtained in Refs.\cite{12} and \cite{13} by the authors of these works.

Third, the phonon spectroscopy in dirty superconductors  or superconductors \cite{1} with a short $\xi$ \cite{4,5} is probably associated with a jump in the chemical potential of pairs in the vicinity of the phase slip surface which is formed so that the main part of the total potential difference across the contact corresponds to it. The latter statement can be naturally extended to the measurements of the phonon structure of PC spectra in the present work.
\section{ENERGY GAP}
The energy gap is manifested in the form of minima of $dV/dI$ on the regions corresponding to increasing excess current (Figs.\ref{Fig2}-\ref{Fig4}). The values of the gap width lie in the interval between 0 and 13.4~$meV$. The form of the $R_D(V)$ curve depends on the HTS region which is the closest to the SN boundary since we assume that the gap is spatially inhomogeneous and is obviously anisotropic. These properties are apparently due to the microscopic nonuniformity of the phase composition of the sample although two gap minima are observed even for contacts of minimum size (see Table \ref{tab1}, contacts Nos.
1 and 5). It can be noted that when there are two gap minima of $dV/dI$, their positions on the $V$-axis are grouped near the values differing by about a factor of two for $T\ll T_c^*$ (contacts Nos.
1 and 4). In the case of a single minimum, its position either corresponds to one of the above- mentioned values (e.g., contact No. 2) or lies between them (contacts Nos. 3 and 4). It could be supposed that in the case of two minima we deal either with two phases with critical temperatures differing also by a factor of two, or with harmonics of the same gap. If, however, the second gap were a harmonic of the first one, the two harmonics would vanish at the same temperature, which is not confirmed in experiments. Besides, the depth of the $dV/dI$ minimum corresponding to the second gap can be either smaller or larger than that of the first minimum (see, for example, curves 1, 1' and 1" in Fig.\ref{Fig3}), and one of them can even disappear. The values of the gap width given in Table \ref{tab1} depend neither on the contact resistance nor on the ratio $R_0/R_N$ reflecting the degree of filling of the contact region with the superconducting phase. It should be noted that for contacts Nos.4 and 5, $T_c^*=20~K$ which is considerably lower than $T_{cbulk}=36~K$. The values of the gaps are also reduced accordingly so that the ratios $2\Delta_1/T_{c1}$ and  $2\Delta_2/T_{c}^*$ remain unchanged. It will be shown in the next section that there is actually a temperature $T_{c1}=(1/2)T_c^*$ for contacts with two gaps, at which the smaller gap vanishes or assumes immeasurably small values.
\begin{figure}[!t]
\includegraphics[width=7cm,angle=0]{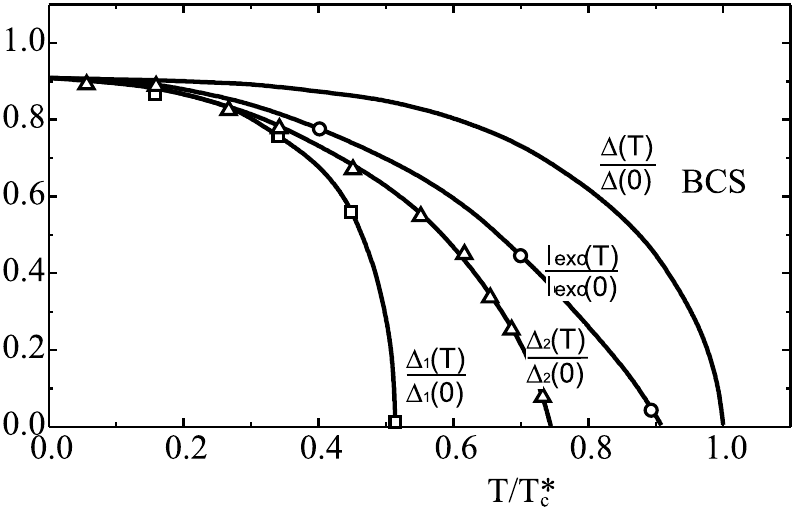}
\caption[]{Temperature dependences of the gaps $\Delta_1$ and $\Delta_2$ and the excess current for contact No.1 in reduced units (the temperature dependence of the gap predicted by the BCS theory is also shown for comparison).}
\label{Fig14}
\end{figure}

\begin{figure}[]
\includegraphics[width=7cm,angle=0]{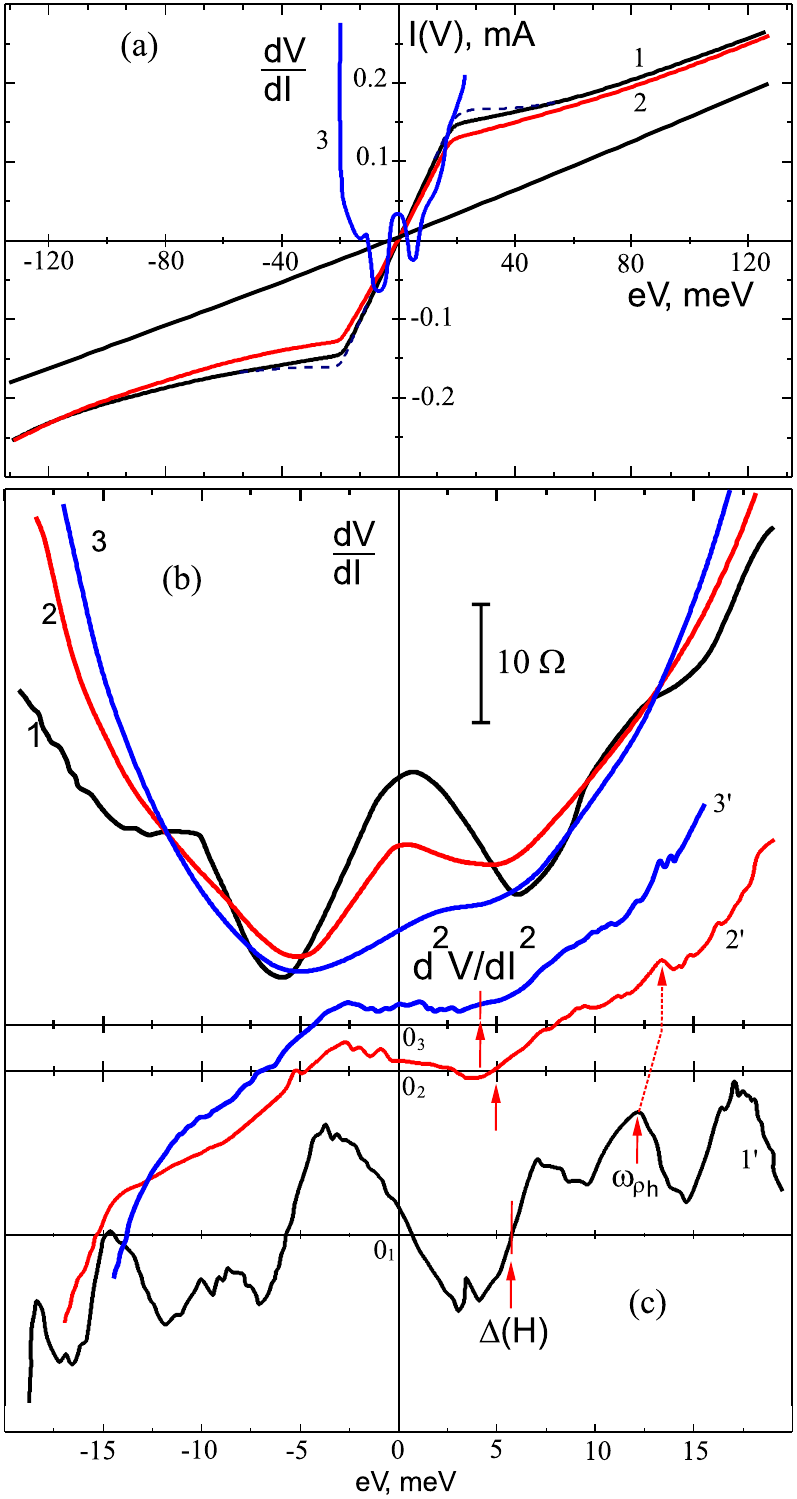}
\caption[]{IVC (a) and their first (b) and second (c) derivatives for contact No.1 in different magnetic fields: curves 1 and 3
in (a) correspond to $H=0$, and curve 2, to $H=32~kOe$; curves 1 and 1' in (b) and (c) correspond to $H=0$; 2 and 2' to $H=21~kOe$; and 3 and 3' to $H=32~kOe$.}
\label{Fig15}
\end{figure}

\begin{figure}[]
\includegraphics[width=7cm,angle=0]{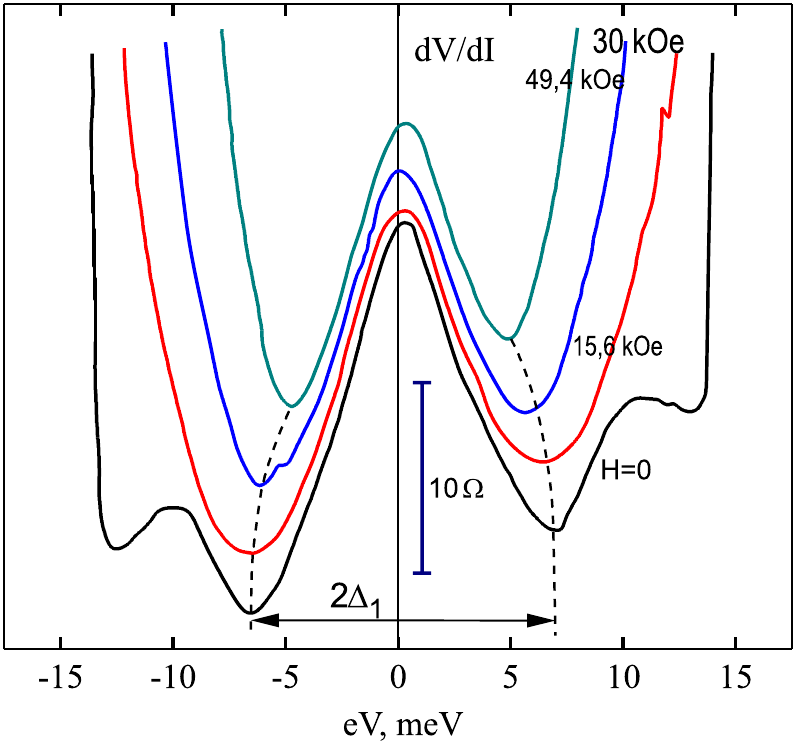}
\caption[]{$dV/dI$ curves for contact No.1 in different magnetic fields in the absence of magnetic flux trapping.}
\label{Fig16}
\end{figure}

\subsection{Temperature dependence}
Figure \ref{Fig10} shows the $dV/dI(V)$ dependencies for contact No.3 at different temperatures. It can be seen that as temperature drops below $T_c^*=31~K$, a minimum appears at zero bias and the derivative $d^2V/dI^2$ changes sign at large bias voltages. By analyzing the characteristics plotted on an extended scale in $V$ and corresponding to smaller temperature differences (Fig. \ref{Fig11}), we note that $dV/dI$ has two minima whose position and intensity depend on temperature. At low temperatures (5, 9.3 and 9.8~$K$), the intensity of the second minimum corresponding to the larger gap is small, and this gap merges with the more intense first minimum on the one hand and is masked by low-frequency phonon singularities which overlap in energy with the gap for this material on the other hand. As the temperature increases, the intensity of the first minimum decreases faster than that of the second minimum, and the two minima become well pronounced in the temperature interval between 11.2 and 15.4~$K$ (Figs. \ref{Fig10} and \ref{Fig11}). It can also be seen that the first minimum approaches zero faster than the second minimum with increasing temperature, and for this reason the second minimum by no means can be treated as a harmonic of the first minimum at temperatures 12-14~$K$. Thus, the gaps $\Delta_1$ and  $\Delta_2$ are of different origin, and the first gap vanishes in the temperature region 15~K ($T_c^*/2$). The
second gap for the contact under investigation can be clearly traced up to 18~$K$. At higher temperatures (see the curves at $T=19.2~K$ and above), only one minimum is left on $dV/dI$ at $V=0$, its intensity being proportional to the superconducting order parameter in the contact region.

Figures \ref{Fig12} and \ref{Fig13} represent the temperature families of $dV/dI$ characteristics for contact No.1.	As the temperature increases starting from 6-7~$K$, it can be seen that the first minimum approaches zero faster than the second one, and their intensities are redistributed. For the given contact, $T_c^*=27.4~K$ and the first gap vanishes at $T_c^*/2=13.7~K$. In the vicinity of this temperature, IVC changes so rapidly in the contact region due to redistribution of current lines (see curves $dV/dI(V)$ for $T$=13.2; 13.8 and 14.1~$K$ in Fig. \ref{Fig12}) that temperature fluctuations do not allow to construct a smooth and reproducible curve $dV/dI(V)$. In order to trace the singularities on $dV/dI(V)$ due to the gap $\Delta_2(T)$ at higher temperature, second derivatives of IVC were recorded (Fig. \ref{Fig13}), the value of $\Delta_2(T)$ on which corresponds to the position of the minimum of $d^2V/dI^2(V)$ from the considerations of "joining" of the $\Delta_2(T)$ dependence in the regions of low and high temperatures. It should be noted that although such a choice is arbitrary, it corresponds to the singularity of $d^2V/dI^2(V)$ near zero with the maximum energy which can still be attributed to the gap. The gap determined in this way obviously vanishes in the temperature region of 19-20~$K$. According to the second derivatives of IVC in the lower part in Fig. \ref{Fig13}, only one structureless minimum remains on the $dV/dI(V)$ dependence at $V$=0 at higher temperatures, which vanishes at $T_c^*=27.4~K$.

The $\Delta_1(T)$ and $\Delta_2(T)$ dependencies constructed for this contact in reduced units are shown in Fig. \ref{Fig14}. They differ from those predicted by the BCS theory. This difference is partially due to the fact that we disregarded the function of thermal smearing of energy distribution for electrons in the normal metal, which broadens the gap singularities on IVC for ScN contacts and shifts them towards lower energies. However, the energy gap differs from the superconducting order parameter (to which the excess current is proportional) even in the region of comparatively low reduced temperatures.

It should be mentioned that the point-contact method of the gap measurement, which is essentially a "current" method, differs from the tunnel method in that the current density in the contact region in the normal and superconducting states is not only very high, but also distributed nonuniformly. Roughly speaking the current chooses the route corresponding to phases with maximum conductivity and the maximum gap.

\subsection{Magnetic field dependence}
Figure \ref{Fig15} shows IVC and their derivatives in different magnetic fields applied in the contact plane. Current-voltage characteristics (Fig. \ref{Fig15}a) indicate that the magnetic field increases the zero-point resistance $R_0$ and reduces the "critical" current $I_c$ corresponding to a transition to the normal differential resistance $R_N$. The excess current also decreases. These changes can be explained by a decrease in the gap width under the action of the field. Indeed, the position of gap minima on $dV/dI$ curves is shifted towards zero, and the minima are broadened (Fig. \ref{Fig15}b).
In addition to the gap singularity at $\Delta_1=6.5~meV$, the $d^2V/dI^2$ curve in zero field exhibits a peculiar structure at $eV=12.5~meV$, which is apparently due to the effect of the gap narrowing by nonequilibrium phonons having a singularity in the density of states at this energy. In spite of the smearing of this singularity it is clearly pronounced even in the field of 21~$kOe$ and is obviously shifted towards higher energies, although the gap singularity for $\Delta_1$ in the same field is shifted in the opposite direction.
\begin{figure}[]
\includegraphics[width=7cm,angle=0]{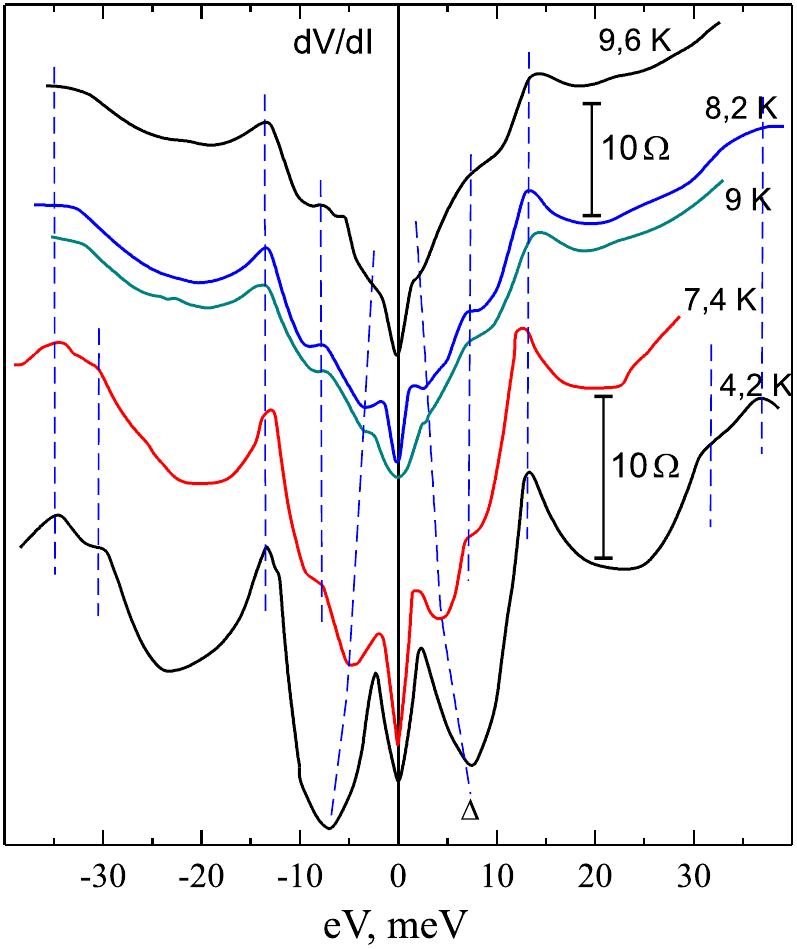}
\caption[]{Initial regions of $dV/dI(V)$ characteristics for contact No.4 at different temperatures.}
\label{Fig17}
\end{figure}

The form of IVC and its derivatives in zero field and their variation with the field are generally determined by the past history, which is apparently due to the magnetic flux trapping in the contact region. This is illustrated in Fig. \ref{Fig16} showing the family of $dV/dI$ curves for the same contact, which was obtained after heating above $T_c^*$ and subsequent cooling in a negligibly small field. By comparing Figs. \ref{Fig15} and \ref{Fig16}, we can conclude that the magnetic flux was obviously trapped in the contact region in the former case, which resulted in the additional smearing of the characteristics. Figure \ref{Fig16} indicates that the effects of the magnetic field on the gap minimum $\Delta_1$ and $\Delta_2$ are qualitatively different. The former minimum is just shifted towards zero practically without broadening (the broadening is observed only with a trapped magnetic flux (Fig. \ref{Fig15})), while the latter minimum vanishes even in comparatively weak fields.
\begin{figure}[]
\includegraphics[width=7cm,angle=0]{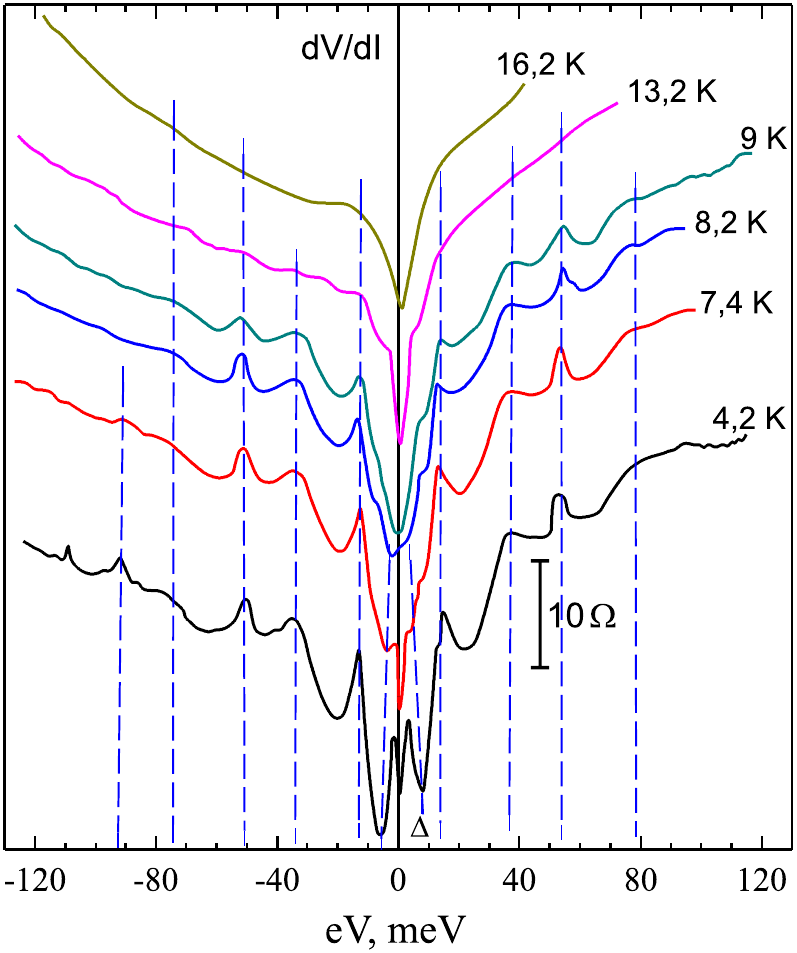}
\caption[]{Phonon structure of the first derivatives of IVC for contact No.4 at different temperatures.}
\label{Fig18}
\end{figure}

Let us suppose that the first minimum is due to elastic reflection of quasiparticles at the boundary between the normal metal and a thin surface layer of HTS with a critical temperature $T_{c1}$ and a gap $\Delta_1$, while the second minimum is caused by the reflection from the boundary between this layer and deeper-lying layers with a critical temperature $T_c^*$ and a gap $\Delta_2$. Figure \ref{Fig6}g presents the model of such a contact if we exclude the normal layer $N_1$ of HTS. As the temperature increases above $T_{c1}$ , the surface layer of HTS becomes normal, and we have a point contact of ScNS type which was investigated in detail in Ref. \cite{14}. The maximum of $dV/dI(0)$ in this case is replaced by a minimum (see, for example, the curves at $T$= 11.2 and 16~$K$ in Fig. \ref{Fig10}). The magnetic field does not affect the tunnel component of the current due to quasiparticles elastically reflected by the first boundary, since they return to the normal electrode in any case. On the contrary, the quasiparticles elastically reflected by the second boundary change their trajectories in the magnetic field so that some of them do not return to the normal electrode, while the current of quasiparticles undergoing an Andreev-type reflection at the second boundary retains its magnitude in the magnetic field since a reflected hole moves in the same curvilinear trajectory as an electron impinging on the second boundary. Consequently, the magnetic field affects the excess current only in proportion to the decrease in the gap $\Delta_2$, while the intensity of the tunnel current component due to this gap can change considerably.

It should be noted that the ballistic mode of quasiparticle transit through a contact, which was presumed in the above analysis, is apparently not essential. The effects of quantum interference may play a significant role in dirty metals if the elastic mean free path $l_i$ and the de Broglie wavelength $h/p_F$ are of the same order of magnitude which is several tens of angstroms for LSCO \cite{15}. For this reason, a magnetic field of the order of $10^4~Oe$ may affect considerably the age current through contacts having diameters- of the order of $d={(\phi_0/H)}^{1/2}=10^{-6}~cm$,  which roughly coincides with the above estimates of the size of high-resistance contacts.
\begin{figure}[]
\includegraphics[width=7cm,angle=0]{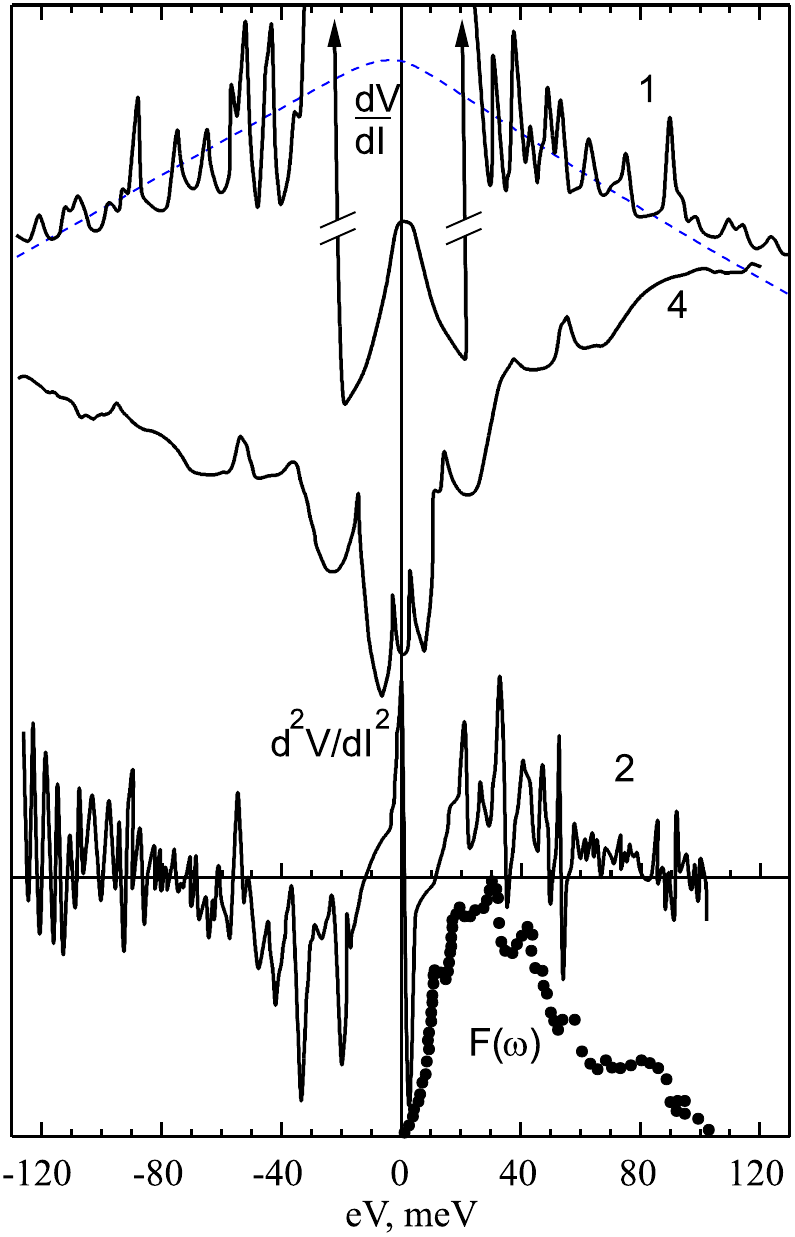}
\caption[]{Phonon structure of IVC derivatives for contacts Nos. 1, 2 and 4. For contact No.2, only the second derivative of IVC is presented: $f(\omega)$ is the density of phonon states according to Ref.\cite{17}.}
\label{Fig19}
\end{figure}
\section{PHONON SINGULARITIES IN PC SPECTRA}
\subsection{Temperature dependences}
It was mentioned above (see Fig. \ref{Fig15}) that the application of a magnetic field shifts some singularities of IVC located in the vicinity of $eV=\Delta$ and manifested more clearly in the second derivatives towards larger bias voltages, although the gap minimum on $dV/dI$ is shifted towards zero as should be expected.
Similar singularities were observed earlier for ScN contacts between $Ta$ and $Cu$ away from the gap minimum and were usually attributed to typical phonon energies of $Ta$ \cite{9}. According to Refs. \cite{4,5}, this mechanism of manifestation of the electron-phonon interaction (EPI) is the principal mechanism for point contacts formed by superconductors with a small coherence length, which is apparently responsible for the manifestation of phonon singularities on IVC of point contacts formed by HTS under investigation. The peculiar feature of HTS is that their gap is anomalously large and can overlap in energy with low-frequency phonon modes. This may lead to a noticeable difference between the true energies of low-frequency phonons and the bias $eV$ at which the phonon peculiarities on IVC derivatives are observed \cite{16}.

The structure of $dV/dI$ characteristics due to the interference between gap and phonon singularities is manifested more clearly in Fig. \ref{Fig17} where the gap minimum is shifted towards zero with increasing temperature, while phonon peaks at 7.2,
13.2~$meV$, etc., practically do not change their position, some of them displaying a small shift towards higher energies. The height of phonon spikes at low temperatures amounts up to 10\% of the value of $R_N$ which is nearly an order of magnitude higher than the intensity of similar spikes on IVC derivatives for $Ta-Cu$ contacts of large diameter \cite{2}. The $dV/dI$ characteristics for contacts No.4 in a wider interval are shown in Fig. \ref{Fig18}. The peaks on $dV/dI(V)$ which we ascribe to EPI are not displaced with temperature, although $dV/dI$ characteristic undergoes large changes in the gap region in the same temperature interval. The phonon singularities are slightly blurred with increasing temperature (up to $T_c^*/2=10~K$) after which their intensity decreases sharply.
\subsection{Comparison with the density of phonon states}
The phonon structure of IVC and its derivatives can be successfully reproduced for a given contact, but varies in detail from contact to contact which is not surprising since the energy gap and the critical temperature also depend on the position of a contact on the surface of HTS. Nevertheless, some general characteristics of this structure can be reproduced for different contacts satisfying the criteria listed in Sec. 2. Figure \ref{Fig19} shows PC spectra for three different contacts (1, 2 and 4). The density of phonon states $F(\omega)$ according to neutron studies \cite{17} is also presented in the figure. The figure shows that at energies higher than $90-100~meV$, the peak intensity drops abruptly. Weaker singularities at higher energies are probably due to harmonics and combinations of different phonon frequencies. The most intense singularities in all spectra lie between 20 and 50~$meV$. It should be noted that there is no phonon structure in comparatively large contacts ($R_N<10~\Omega$), which is probably due to the formation of a large number of PSC along current lines in the contact region.

A comparison of PC spectra with the phonon density of states indicates that the singularities under consideration are caused by EPI in a given material. The difference in the exact positions of the peaks on the energy axis for different contacts can be either due to the fact that we measured local characteristics of EPI function which may vary considerably from point to point, or because the positions of the peaks on the energy axis do not always correspond to phonon frequencies exactly, but just have a tendency to group near them. More detailed investigations on high-quality single crystals will probably permit a complete reconstruction of the set of characteristic energies of phonons and their relative contribution to the EPI function depending on the contact axis orientation relative to the crystallographic axes and other parameters.

We believe that the results presented in this work lead to the conclusion that inelastic processes of carrier scattering in $La_{1.8}Sr_{0.2}CuO_4$ occur predominantly on phonons. This means that the electron-phonon iteration in this material is decisive and in all probability forms the basis of Cooper pairing.
\section{CONCLUSIONS}

\begin{enumerate}
\item {A point contact with a size of the order of several tens of angstroms makes it possible to gain information about new high-temperature superconductors, which cannot be obtained in principle by using other methods.}

\item {Two gaps often observed in PC spectra belong to the same microscopic region of the material in the vicinity of the contact. Two gap singularities on the $dV/dI$ characteristics correspond to different phases having different critical temperatures.}

\item {The point-contact method makes it possible to measure the maximum gap (or gaps) in the volume under investigation since it corresponds to the "easiest" paths of electric current.}

\item {The maximum gap has a width of 5-11~$kT_c$, which corresponds to the superconductivity with
a strong EPI.}

\item {Phase slip centers (lines or surfaces) existing in the contact region possibly play an important role in the gap and phonon spectroscopy.}

\item {The electron-phonon interaction is decisive for $La_{1.8}Sr_{0.2}CuO_4$ high-$T_c$ superconductor. The singularities of IVC derivatives ascribed by us to EPI are observed only for bias voltages corresponding to phonon energies. The maximum intensity is observed for the singularities lying near the phonon frequencies with the maximum density of states. There are low-frequency phonon modes which overlap in energy with the gap in the quasiparticle excitation spectrum of a superconductor.}

\end{enumerate}
The authors are grateful to B.I.~Verkin for his attention and encouragement and also to N.B.~Brandt and V.V.~Moshchalkov who initiated the present investigation.
\section{NOTATION}
Here $d$ is the contact diameter, $l_i$ and $l_{\varepsilon}$ the elastic and inelastic mean free paths, $\Lambda_{\varepsilon}$ the energy relaxation length, $l_E$ the electric field penetration depth, $\Delta$ the energy gap, $\xi$ the coherence length,
$I_{exc}$ the excess current, and $\rho$ the resistivity of the material.


\begin{thebibliography}{}


\bibitem{1}	I. K. Yanson, N. L. Bobrov, L. F. Rybal'chenko, and V.~V. Fisun, \href{http://fntr.ilt.kharkov.ua/fnt/pdf/9/9-11/f09-1155r.pdf}{Fiz. Nizk. Temp.} \textbf{9}, 1155 (1983) [Sov. J. Low Temp. Phys. \textbf{9}, 596 (1983)], \href{https://arxiv.org/pdf/1604.07067.pdf}{arXiv:1604.07067}.
\bibitem{2}	I. K. Yanson, V. V. Fisun, N. L. Bobrov, and L. F. Rybal'chenko, \href{http://www.jetpletters.ac.ru/ps/141/article_2443.pdf}{Pis'ma Zh. Eksp. Teor. Fiz.} \textbf{45}, 425 (1987) [\href{http://www.jetpletters.ac.ru/ps/1244/article_18813.pdf}{JETP Lett.} \textbf{45}, 543 (1987)], \href{https://arxiv.org/pdf/1602.04356.pdf}{arXiv:1602.04356}.
\bibitem{3}	I. O. Kulik, A. N. Omel'yanchuk, and I. Beloborodko, in: Proc. of the \href{https://www.amazon.com/Proceedings-Soviet-Italian-Symposium-Weak-Superconductivity/dp/9971505045}{Second Soviet-Italian Symposium on Weak Superconductivity}, May 1987, Naples, Italy, A. Barone and A. Larkin (eds.), World Scientific, Singapore-New Jersey-Hong Kong (1987).
\bibitem{4}	I. K. Yanson, N. L. Bobrov, L. F. Rybal'chenko, and V. V. Fisun, Fiz. Tverd. Tela (Leningrad) \textbf{27}, 1795 (1985) [Sov. Phys. Solid State \textbf{27}, 1076 (1985)].
\bibitem{5}	N. L. Bobrov, L. F. Rybal'chenko, M. A. Obolenskii, and V. V. Fisun, \href{http://fntr.ilt.kharkov.ua/fnt/pdf/11/11-9/f11-0925r.pdf}{Fiz. Nizk. Temp.} \textbf{11}, 925 (1985) [Sov. J. Low Temp. Phys. \textbf{11}, 510 (1985)], \href{https://arxiv.org/pdf/1603.02598.pdf}{arXiv:1603.02598}.
\bibitem{6}	I. K. Yanson, L. F. Rybal'chenko, V. V. Fisun,  N. L. Bobrov, M. A. Obolenskii,
N. B. Brandt, V. V. Moshchalkov, Yu. D. Tret'yakov, A. R. Kaul, and I. E. Graboi, \href{http://fntr.ilt.kharkov.ua/fnt/pdf/13/13-5/f13-0557r.pdf}{Fiz. Nizk.Temp.} \textbf{13}, 557 (1987) [Sov. J. Low Temp. Phys. \textbf{13}, 315 (1987)), \href{https://arxiv.org/pdf/1701.01982.pdf}{arXiv:1701.01982}.
\bibitem{7}	G. Deutscher and K. A. Muller, \href{https://doi.org/10.1103/PhysRevLett.59.1745}{Phys. Rev. Lett.} \textbf{59}, 1745 (1987).
\bibitem{8}	E. R. Moog, M. E. Hawley, K. E. Gray, et al., \href{http://link.springer.com/article/10.1007/BF00116870}{J. Low Temp.Phys.} \textbf{71}, 393 (1988).
\bibitem{9}	I. K. Yanson, N. L. Bobrov, L. F. Rybal1chenko, and V. V. Fisun, \href{http://fntr.ilt.kharkov.ua/fnt/pdf/13/13-11/f13-1123r.pdf}{Fiz. Nizk. Temp.} \textbf{13}, 1123 (1987) [Sov. J. Low Temp. Phys. \textbf{13}, 635 (1987)], \href{https://arxiv.org/pdf/1512.03917.pdf}{arXiv:1512.03917}.
\bibitem{10}	V. M. Dmitriev, I. V. Zolochevskii, and E. V. Khristenko,\href{http://fntr.ilt.kharkov.ua/fnt/pdf/14/14-2/f14-0134r.pdf}{Fiz. Nizk. Temp} , \textbf{14}, 134 (1988) [Sov. J. Low Temp. Phys. \textbf{14}, 73 (1988).
\bibitem{11}	I. K. Yanson, L. F. Rybal'chenko, N. L. Bobrov, and V. V. Fisun, \href{http://fntr.ilt.kharkov.ua/fnt/pdf/12/12-5/f12-0552r.pdf}{Fiz. Nizk. Temp.} \textbf{12}, 552 (1986) [Sov. J. Low Temp.Phys. \textbf{12}, 313 (1986)), \href{https://arxiv.org/pdf/1512.00684.pdf}{arXiv:1512.00684}.
\bibitem{12}	A. Hahn, \href{https://doi.org/10.1103/PhysRevB.31.2816}{Phys. Rev. B}\textbf{31}, 2816 (1985).
\bibitem{13}	A. Hahn and K. Humpfner, in: Proc. 18-th Int. Conf. on Low-temp. Phys., Kyoto, 1987, \href{http://iopscience.iop.org/article/10.7567/JJAPS.26S3.1599/pdf}{Jpn. J. Appl. Phys.} \textbf{26}, Suppl. 26-3, 1599 (1987).
\bibitem{14}	P. C. van Son, H. Kempen, and P. Wyder, \href{https://doi.org/10.1103/PhysRevLett.59.2226}{Phys. Rev. Lett.} \textbf{59}, 2226 (1987).
\bibitem{15}	L. N. Gor'kov and N. B. Kopnin, \href{http://ufn.ru/ufn88/ufn88_9/Russian/r889d.pdf}{Usp. Fiz. Nauk} \textbf{156}, 117 (1988) [Sov. Phys. Usp. \textbf{31}, 850 (1988)].
\bibitem{16}	G. V. Kamarchuk, A. V. Khotkevich, and I. K. Yanson, Fiz.Tverd. Tela (Leningrad) \textbf{28}, 455 (1986) [Sov. Phys. Solid State \textbf{28}, 254 (1986)].
\bibitem{17}	P. P. Parshin, M. G. Zemlyanov, N. N. Chernoplekov, et al., in: Superconductivity: Physics, Chemistry, Engineering, V. I. Ozhogin (ed.) [in Russian], Atomic Energy Institute, Issue 2, Moscow (1988).


\end{thebibliography}
\end{document}